\documentclass[final,3p,times,twocolumn]{elsarticle-mod}

\usepackage{amssymb}

\usepackage[hyphens]{url}

\usepackage{amsmath}

\usepackage{color}
\usepackage{xtab}
\usepackage{overpic}
\usepackage[para]{threeparttable}
\usepackage{etoolbox}
\usepackage{pict2e}
\usepackage{fancyhdr}
\newcommand{\MARK}[1]{#1}
\newcommand{\MARKI}[1]{#1}
\newcommand{\MARKII}[1]{#1}
\newcommand{\MARKIII}[1]{#1}
\newcommand{\MARKIV}[1]{#1}

\urldef\cplusplusurl\url{https://en.wikipedia.org/wiki/C%2B%2B14}
\appto\TPTnoteSettings{\footnotesize}

\journal{NSS Space Settlement Journal}

\begin{document}

\begin{frontmatter}



\title{\MARK{Natural illumination} solution for \MARK{rotating} space \MARKIV{settlements}}


\author[FMI]{Pekka Janhunen\corref{cor1}}
\ead{pekka.janhunen@fmi.fi}
\ead[url]{http://www.electric-sailing.fi}

\address[FMI]{Finnish Meteorological Institute, Helsinki, Finland}
\cortext[cor1]{Corresponding author}

\begin{abstract}
  Cylindrical kilometre-scale artificial gravity space \MARKIV{settlements} were
  proposed by Gerard O'Neill in the 1970s. The early concept had
  two oppositely rotating cylinders and moving mirrors to simulate the
  diurnal cycle. Later, the Kalpana One concept exhibited passively stable
  rotation and no large moving parts. Here we propose and analyse a
  specific light transfer solution for Kalpana One type
  \MARKIV{settlements}. \MARK{Our proposed solution is technically reliable because it avoids
    large moving parts that could be single failure points}. The scheme has an array of cylindrical paraboloid
  concentrators in the outer wall and semi-toroidal reflectors at the
  equator which distribute the concentrated sunlight onto the living surface. The
  living cylinder is divided into a number of $\varphi$-sections
  (valleys) that are in different phases of the diurnal and seasonal cycles. To reduce the mass of nitrogen needed,
  a shallow atmosphere is used which is contained by
  a pressure-tight transparent roof. The only moving parts needed are local blinders
  installed below the roof of each valley. We also find
  that \MARKIV{settlements} of this class have a natural location at the equator
  where one can build multi-storey urban blocks. The location is optimal from the
  mass distribution (rotational stability) point of view. If maximally built, the amount of urban
  floorspace per person becomes large, up to 25,000 m$^2$, which is an order of magnitude
  larger than the food-producing rural biosphere area per
  person. Large urban floorspace area per person may increase the
  material standard of living much beyond Earth while increasing the total
  mass per person relatively little.
\end{abstract}


\end{frontmatter}



\section*{Nomenclature}
\nobreak\noindent
\begin{tabular}{ll}
\MARKII{$A_\perp$}    & \MARKII{Cross-sectional area} \\
au     & Astronomical unit, 149\,597\,871\,km \\
\MARKII{$F_y$}    & \MARKII{Wall tension force} \\
$g$    & Acceleration due to gravity, 9.81 m/s$^2$ \\
\MARKII{$h$}    & \MARKII{Wall thickness} \\
$K$    & Concentration factor \\
\MARK{$L_z$}  & \MARK{\MARKIV{Settlement} length} \\
\MARKII{$m$}  & \MARKII{Mass} \\
$p$    & Half slitwidth (\emph{semilatus rectum}) of concentrator \\
$R$    & \MARK{Radius of the living wall} \\
\MARK{$x,y,z$} & \MARK{Cartesian coordinates} \\
\MARKII{$\rho$}  & \MARKII{Radial cylindrical coordinate, $\rho=\sqrt{x^2+y^2}$} \\
\MARKII{$\varphi$} & \MARKII{Angular cylindrical coordinate, $\varphi=\mathrm{atan}(y/x)$} \\
\MARK{$\rho_w$} & \MARK{Wall mass density (kg/m$^3$)} \\
\MARKII{$\sigma$} & \MARKII{Wall tension (Pa)} \\
\end{tabular}


\section{Introduction}

Gerard O'Neill was among the first to propose that people could live
on the insides of kilometre-scale spinning cylindrical \MARKIV{settlements},
located in space and constructed from asteroid or lunar materials
\citep{ONeill1974,ONeill1977}. O'Neill's original concept had two cylinders
rotating in opposite directions and mechanically connected to each
other to make a system with almost zero total angular momentum so that
it can be turned to track the Sun. The drawback of such design is the
existence of large rotary joints and moving mirrors that are
potential sources of single-point failures. Later, the Kalpana One
model was proposed \citep{KalpanaOne}, which consisted of a single
cylinder whose axis of rotation is perpendicular to the orbital
plane, thus eliminating much of the mechanical complexity.

It has been pointed out \citep{ONeill1974,ONeill1977,KalpanaOne} that
unlike Moon and Mars, artificial rotating \MARKIV{settlements} are able to provide
an earthlike $1 g$ gravity environment for the inhabitants which
ensures that children grow as strong as on Earth so that they are free
to visit or move back to Earth as adults if they wish. Also, in the
long run, there is enough small body material in the asteroid belt,
and even more in the Kuiper belt and beyond, that the total living
area of \MARKIV{settlements} could eventually exceed the surface area of Earth by
a large factor.  This is understandable because a \MARKIV{settlement} needs only
$\sim 10^4$ kg/m$^2$ of radiation shielding mass per sunlit living
wall area, while an earthlike planet needs a million times more. The
\MARKIV{settlement} technology allows one to build $1 g$ living space in the solar
system by using the mass per area equivalent to Earth's atmosphere
only, rather than the entire planet.

On Moon and Mars it is possible to build radiation shielded and
pressurised living space underground e.g.~in existing caves or lava
tunnels, with low expenditure of processed structural materials such
as steel. In contrast, rotating cylindrical \MARKIV{settlements} in free space
need more structural materials and therefore they depend on the
existence of scaled-up asteroid mining, space manufacturing and
low-cost in-space propulsion such as electric sail \citep{RSIpaper}
for moving the materials and equipment. However, there seems to be no
reason why asteroid mining and space manufacturing, once started,
could not scale up exponentially.

While O'Neill's groundbreaking idea of rotating cylindrical ``inverted
planets'' received significant public attention and gave rise to
conferences and workshops where different \MARKIV{settlement} geometries were
considered (for a review, see \citet{Marotta2016}), the peer-reviewed
literature on cylindrical rotating \MARKIV{settlements} is rather small. In
addition to the references already mentioned, an example of this
literature is a study of the ultimate limit in \MARKIV{settlement} radius with
advanced materials \citep{McKendree1996}.

In this paper we study a class of Kalpana One type \MARKIV{settlements}
\citep{KalpanaOne}, but with a specific arrangement for transferring
sunlight to mimic earthly diurnal and seasonal cycles without large
moving parts. The O'Neill and Kalpana One cylinders were assumed to be
filled with breathable atmosphere, but here we assume a relatively
shallow (50 m) atmosphere under a transparent pressure-holding roof in
order to reduce the amount of nitrogen required.  Nitrogen is a
relatively rare element on asteroids and on the Moon. Although humans
could also survive in a reduced pressure oxygen atmosphere, such
atmosphere would introduce an elevated risk of fire.  Furthermore,
flying insects and birds would have challenges in keeping aloft in a
reduced pressure pure oxygen atmosphere, because the mass density of
such atmosphere would be much lower than at sea level on Earth. Flying
animals play important roles in the internal ecosystem, for example
flying insects are responsible for pollination of fruits and
berries. In addition to being more affordable in terms of nitrogen,
elimination of a cylinder-filling pressurised atmosphere reduces the
tensile strength requirement of the walls significantly.

For the location of the \MARKIV{settlement}, we assume a 1\,au orbit,
for example the Earth-Moon L4 or L5 Lagrange point or
the Earth-Sun L4 or L5 Lagrange point. More general orbits would be
possible, but their study is not in the scope of this paper. We assume $1 g$
artificial gravity, 1\,bar atmospheric pressure and earthlike
radiation protection provided by $10^4$ kg/m$^2$ thick walls.

The focus of the paper is a future where -- we assume --
common engineering materials such as steel, aluminium and carbon fibre
composites are available in scaled-up abundance from asteroid or lunar
resources, and where the question of the preferred \MARKIV{settlement} size (the
unit size of large-scale solar system \MARKIV{settling}) is
relevant.  Most inhabitants likely prefer a large \MARKIV{settlement} if given the
choice. However, the tensile strength requirement of the \MARKIV{settlement} wall
grows linearly with the \MARKIV{settlement} radius. Thus for a larger \MARKIV{settlement}, a
larger fraction of the wall mass must be structural material instead
of biospheric payload like soil, vegetation and people. On the other
hand, the radiation shielding requirement sets a minimum areal mass
density for the wall which we assume to be $10^4$ kg/m$^2$. Then there
exists a \MARKIV{settlement} size -- the sweet spot -- where the wall
thickness driven by structural requirements also provides the right
amount of radiation shielding. Smaller \MARKIV{settlements} would need equally thick walls
because of the radiation shielding requirement and thus they would
have the same cost per inhabitant, if cost is measured by the needed
asteroid or lunar mass. Larger \MARKIV{settlements}, on the other hand, would need
thicker walls because of the structural requirements, and thus higher
mass expenditure per person.

\rfoot{\textit{NSS Space Settlement Journal}}
\pagestyle{fancy}

\MARKI{Concerning \MARKIV{settlement} scale,} in this paper we adopt 5 km \MARKIV{settlement} radius as representing \MARKI{
  an order of magnitude relevant for long-term
  future}, \MARKII{see Appendix A for wall tensile calculations.}
There \MARK{exists} the timely question of
\MARK{which} size of \MARK{\MARKIV{settlements}} one should build \MARK{in near
  future} \citep{EasierWay}, but \MARK{such} question is
\MARK{outside} the scope of the present paper. \MARK{Indeed, we stress that}
our choice of 5 km radius should \emph{not} be
taken as a recommendation for the first \MARKIV{settlement}. Our lighting solution
is independent of the \MARKIV{settlement} size, however, and thus \MARK{it could}
be applicable \MARK{also for smaller \MARKIV{settlements} in near future}.

An overarching motivation of the paper is to find a long-term reliable \MARKIV{settlement}
architecture which enables natural sunlight and earthlike (and
configurable) diurnal and seasonal illumination cycles. We take the
long-term reliability requirement to imply that large moving parts and
single failure points must be avoided and that the rotation must be
passively stable. As we will find out below, a lighting solution and the
associated \MARKIV{settlement} architecture exists that satisfies the requirements.

The structure of the paper is as follows. We describe the \MARKIV{settlement}
geometry, present a ray-tracing simulation to calculate how sunlight
photons are distributed inside the \MARKIV{settlement}, then discuss the placement
of urban blocks\footnote{\MARKII{By urban blocks we refer to multi-storey
  floorspace that has no natural illumination.}}. We find that the naturally buildable urban floor area
per inhabitant is large, which tends to promote the
standard of living. We close the paper by discussion, summary and
conclusions. The focus of the paper is on the lighting solution;
other features of the \MARKIV{settlement} are considered as needed. The software
used to compute the results is available at \citep{raytrace}.

\MARK{

\section{Light channel}

Directional light such as sunlight can be concentrated e.g.~by a
parabolic reflector or lens. If the concentrated light is directed
into a box (light channel) whose inner walls are perfectly reflecting,
the concentrated light fills the entire box because photons can only
exit through the same slit where they entered (Figure
\ref{fig:lightbox}).  If the concentration ratio is increased, the
light intensity inside the box increases proportionally to it, until
reaching the surface brightness of the Sun.

In reality the light intensity is lower because the walls are not
ideally reflecting, and the light intensity also depends on the size
and shape of the box. For our \MARKIV{settlement} application, we prefer a light
intensity in the light channel which is a few times solar. This is
because for safety reasons we want to impose a constraint that if the
light channel contains some dark object or dark region (which might be
necessary during servicing, for example), the local temperature must
not go too high.

\begin{figure}[htbp]
\centering
\begin{overpic}[width=\columnwidth]{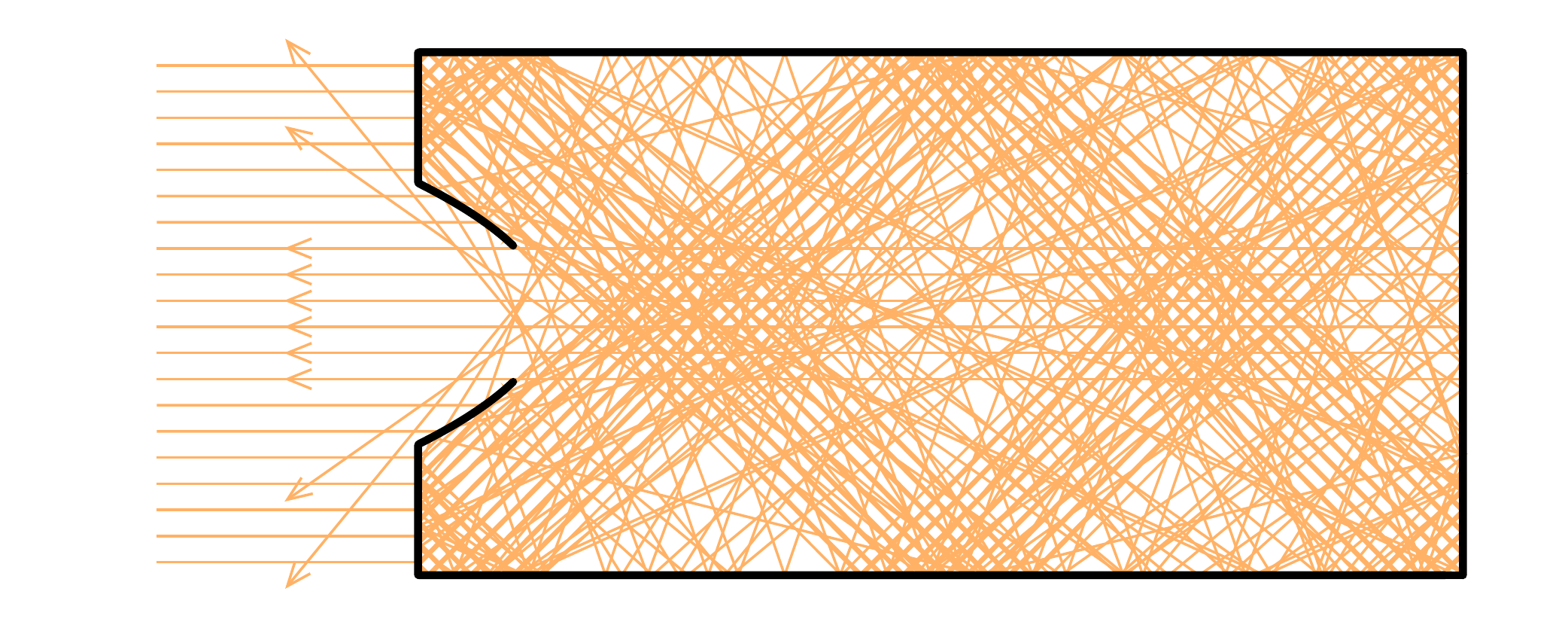}\put(4,34){a)}\end{overpic}
\begin{overpic}[width=\columnwidth]{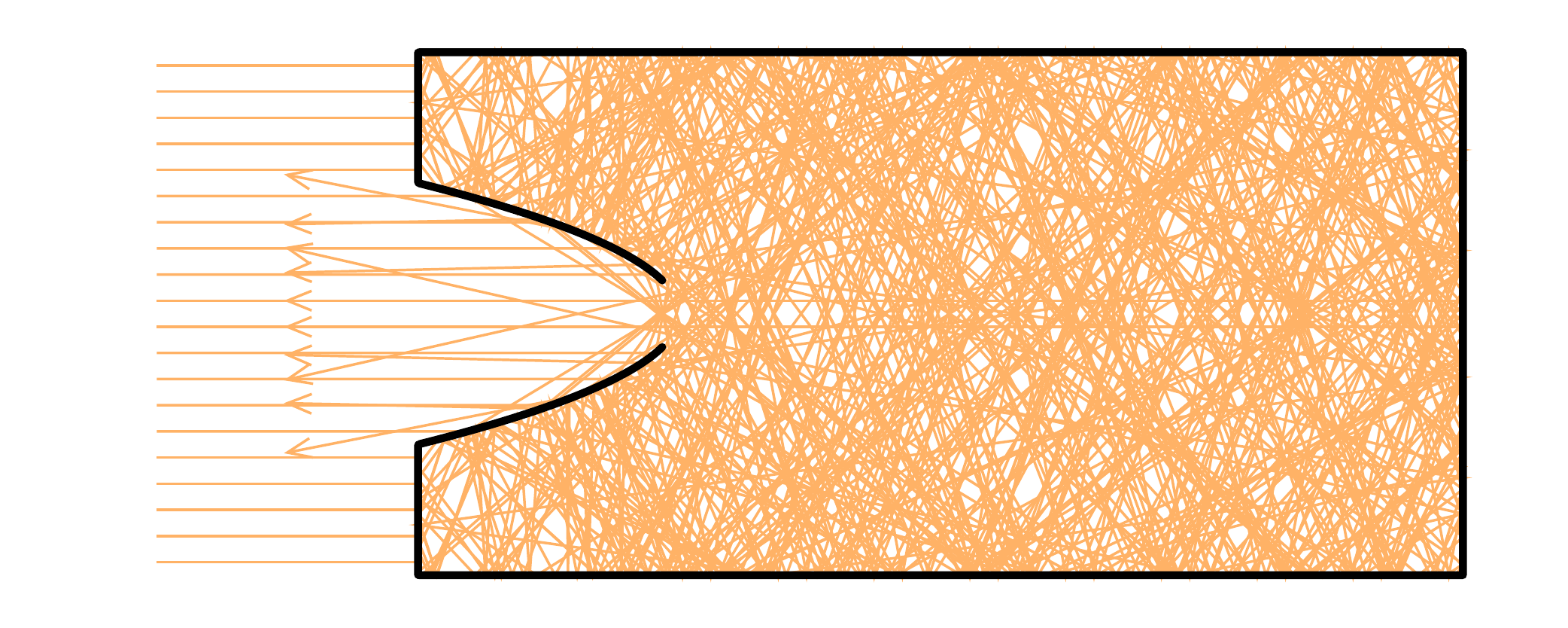}\put(4,34){b)}\end{overpic}
\caption{
\MARK{When sunlight is concentrated into a light channel by a parabolic
concentrator, photons can only escape through the same slit where they
entered. Light intensity inside the box in case (b) is higher than in case (a) because the
concentration ratio is higher.}
}
\label{fig:lightbox}
\end{figure}

Sunlight into the \MARKIV{settlement} is concentrated by fixed cylindrical
paraboloid concentrators into the light channel
(Fig.~\ref{fig:blockdiag}).  The light channel has a geometric shape
that distributes light around the \MARKIV{settlement}, in particular delivering
light also to the antisunward side. The \MARKIV{cylindrical} rural wall surface (the living
cylinder \MARKIV{where the biosphere exists}) taps its illumination from the adjacent light channel
through windows that form the local roof of the rural surface. Local
controllable blinders are used in the windows to implement desired
diurnal and annual light cycles, in each $\varphi$-zone of the rural
wall separately. When closed, the blinders reflect light back to the
light channel where it remains usable for other $\varphi$-zones whose blinders
are open.

\begin{figure}[ht!]
\centering
\fbox{\includegraphics[width=0.35\columnwidth]{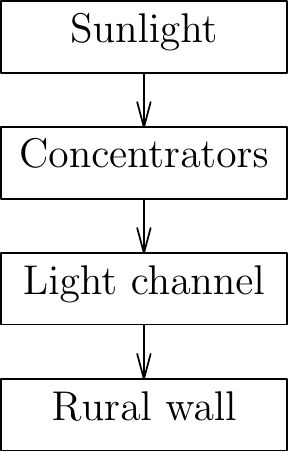}}
\caption{\MARK{Path of sunlight onto the \MARKIV{settlement}'s rural wall
    living area.}
\label{fig:blockdiag}}
\end{figure}


}

\section{\MARKIV{Settlement} geometry}

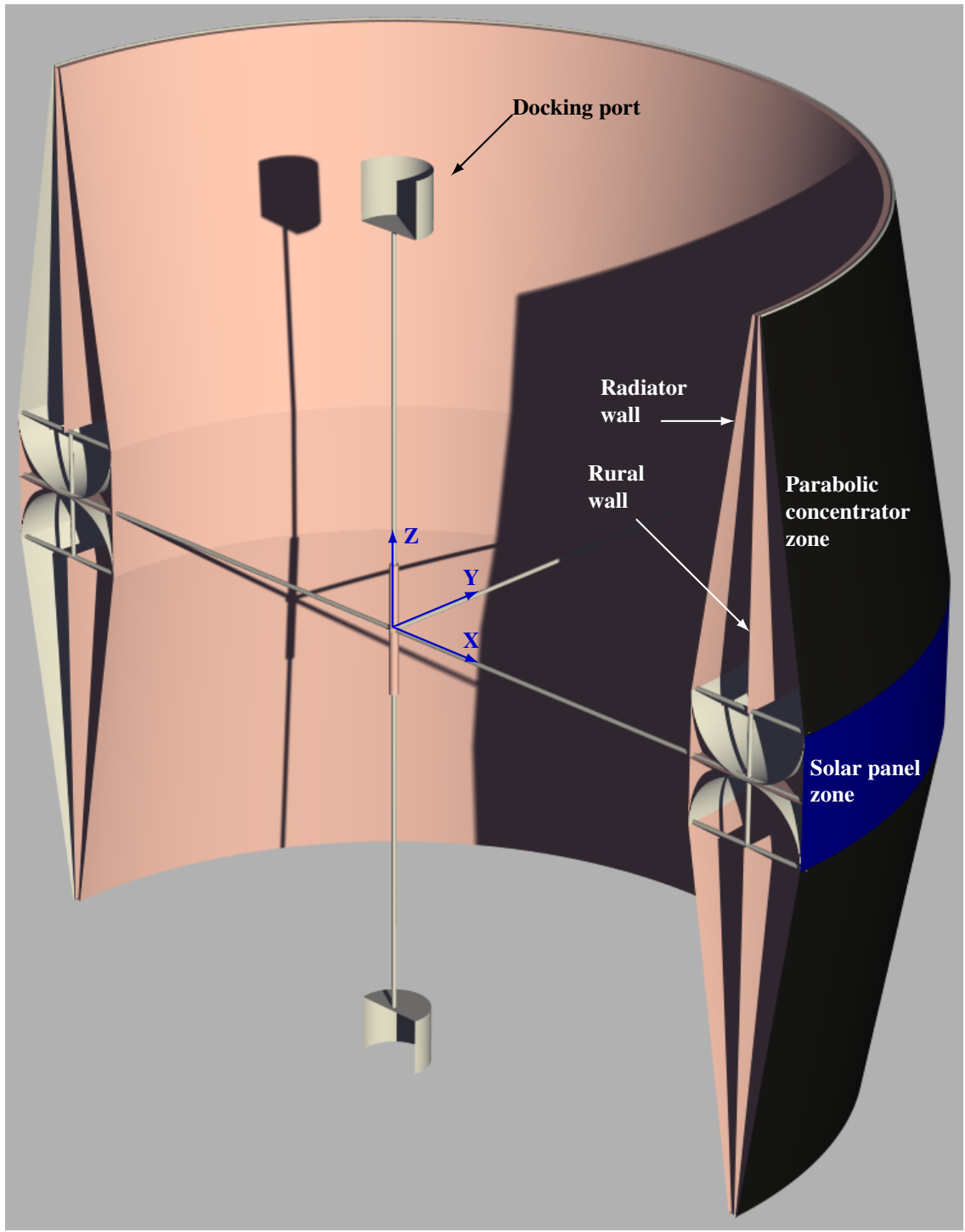
\begin{figure*}[htbp]
\centering
\begin{overpic}[width=1.9\columnwidth]{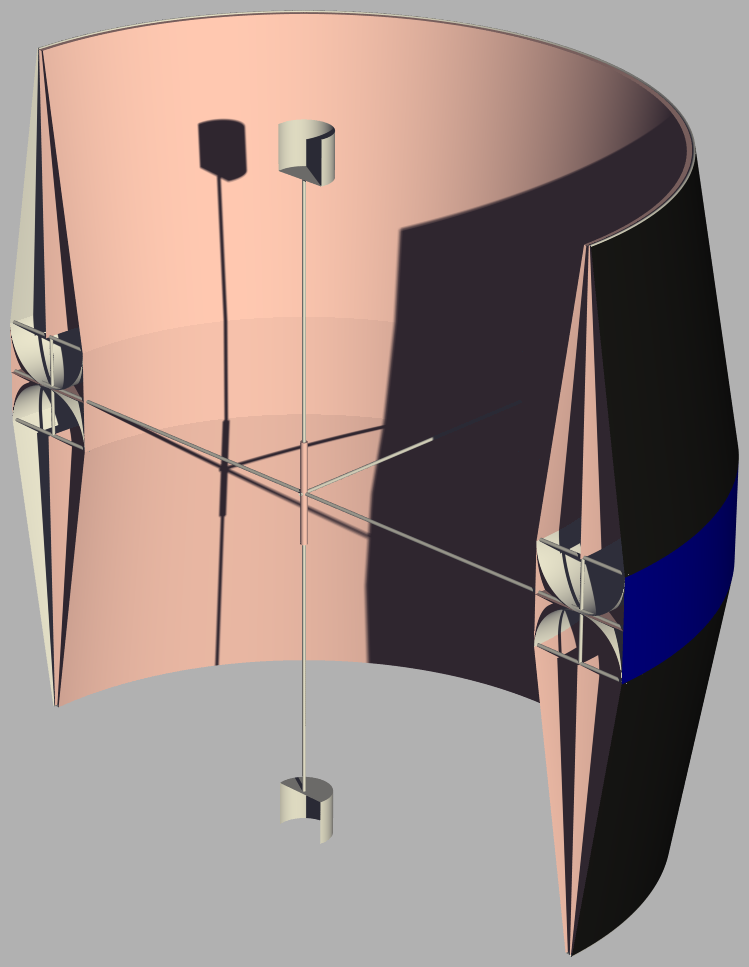}
  \thicklines\bfseries
  \put(41,91){\vector(-1,-1){5}}
  \put(41,91){Docking port}
  \color{white}
  \put(64,36){\begin{tabular}{l}Solar panel\\zone\end{tabular}}
  \put(62,58){\begin{tabular}{l}Parabolic\\concentrator\\zone\end{tabular}}
  \put(51,58){\vector(1,-1){9.2}}
  \put(46,60){\begin{tabular}{l}Rural\\wall\end{tabular}}
  \put(53,66){\vector(1,0){6}}
  \put(47,67){\begin{tabular}{l}Radiator\\wall\end{tabular}}
  \color[rgb]{0,0,0.8}
  \newcommand{\x}{31.3}
  \newcommand{\y}{49.2}
  \put(\x,\y){\vector(0,1){8}}
  \put(\x,\y){\vector(2,0.84){7}}
  \put(\x,\y){\vector(2,-0.84){7}}
  \put(32.2,56){Z}
  \put(37,47.5){X}
  \put(37,52.6){Y}
\end{overpic}
\caption{Cut view of the cylindrical space \MARKIV{settlement}.}
\label{fig:schematic3D}
\end{figure*}

A cut 3-D rendering of the \MARKIV{settlement} is shown in
Fig.~\ref{fig:schematic3D}. The overall shape is a cylinder without
endcaps which rotates so that the spin axis is orthogonal to the plane
of the \MARKIV{settlement}'s heliocentric orbit, i.e.~that the spin axis is
perpendicular to the Sun direction. (The illumination direction is
different in Fig.~\ref{fig:schematic3D} to aid visualisation of the shape.)
The \MARKIV{settlement} is divided into identical southern and
northern cylinders separated by an equatorial region. The cylinder has three walls. The
outermost wall has an array of cylindrical paraboloid concentrators, shown as
black in Fig.~\ref{fig:schematic3D}, whose purpose is to trap sunlight
into the light channel beneath. A concentrator centred at $z=z_1$ is the
parametric surface
\begin{equation}
\rho(\varphi,z) = R + \frac{(z-z_1)^2}{2p}, \varphi \in [0,2\pi),
\vert z-z_1 \vert \in [p,Kp]
\end{equation}
where $p$ is the half-width of the concentrator slit (the \emph{semilatus
rectum} parameter of the parabola) and $K=20$ is the concentration ratio.

The middle wall is the \MARKIV{cylindrical} rural \MARKIV{wall} whose radius
\MARK{is taken} to be 5 km. The inner wall reflects the concentrated sunlight
that arrives from the equatorial semi-toroidal reflector\footnote{\MARKII{A
  cylindrical corner reflector made of two conical surfaces was also
  tried, but seemed to distribute light to the backside somewhat less
  efficiently than the toroidal version.}} towards the
rural \MARKIV{wall}.

The biosphere of the \MARKIV{settlement} resides on the inner surface of the
\MARKIV{cylindrical} rural \MARKIV{wall}. \MARKI{About} 50 m \MARKI{above} the living surface there is a
pressure-tight transparent \MARKI{roof} equipped with
0--100\,\% adjustable blinders. The nitrogen-oxygen atmosphere has
not more than 50 m thickness to limit the mass of nitrogen needed. The biosphere
is divided into 20 compartments called valleys in the $\varphi$
direction. The diurnal and seasonal cycles are simulated by
controlling the blinders in each valley separately. The neighbouring
valleys have approximately opposite diurnal and seasonal phases so that the
total dissipated power stays approximately constant. The yearly
average insolation is 100 W/m$^2$ at the high latitude end of the
valley and increases linearly to 160 W/m$^2$ towards the equatorial
end. The values 100 and 160 W/m$^2$ correspond to yearly average
insolation in Helsinki and in southern France, respectively.  We
assume average albedo of the biosphere of 0.2.

During the simulated night of the valley, the blinders of the valley
are closed.  Closed blinders reflect photons back into the light
channel so that the photons benefit other valleys whose blinders are
open.  During daytime the blinders are opened partially to simulate
the wanted light level.

The equatorial part of the \MARKIV{settlement} has semi-toroidal reflectors whose
purpose is to turn the flux of concentrated sunlight coming from the paraboloid
reflectors by approximately 180$^{\circ}$. The equatorial region has a
belt of solar panels shown as dark blue in Fig.~\ref{fig:schematic3D}
to provide electric power for the \MARKIV{settlement}.

The interior of the cylinder has two docking ports for external
spacecraft. Each docking port is a cylinder with one open end.  All
parts rotate with the same angular speed so the cylinder walls of the
docking ports experience a small artificial gravity of 0.074 g. An
arriving spacecraft flies inside the cylinder and sets its landing
gear wheels into rotation that matches the rotational velocity of the
docking port wall. Then it moves slowly to the rotating wall, perhaps
uses some amount of magnetic attachment to prevent bouncing, and
applies braking to the wheels. Due to the braking, the spacecraft
gradually starts to co-move with the wall and the centrifugal force
increases, ensuring that the wheels stay firmly on the wall even
without magnetic attachment. Finally the spacecraft comes to rest with
respect to the rotating wall and experiences the same centrifugal
acceleration (artificial gravity) than the wall. The system can be
thought of as a cylindrical landing strip where wheeled craft of
different sizes may land.

The edge of the cylinder contains one or more inclined ramps.  A
departing spacecraft rolls over the edge of the landing port cylinder
along a ramp, like a wheeled aeroplane that starts to accelerate
downhill. After exiting the ramp and unless propulsively changed, the
spacecraft follows a straight path and moves with the rotational
velocity of the docking port wall. The docking port is placed outside
the $z$ range of other structures of the \MARKIV{settlement} so that the departing
spacecraft does not collide any obstacle.

The arrival and departure procedures sketched in the previous two
paragraphs are such that the spacecraft need only low-thrust
proopulsion. Impulsive chemical propulsion is not mandatory, which is
good from the safety point of view. If chemical propulsion is
nevertheless used for some external reason and if an accidental
explosion occurs in the docking port, the open cylinder shape of the
docking port tends to direct the explosive energy away from the
\MARKIV{settlement}. Existence of two docking ports enables access to the \MARKIV{settlement}
through the other port while a damaged port is repaired.

The docking ports are connected with a central hub by train tubes that
carry passengers and cargo. The central hub contains a recreational
low-g space. It is connected with the cylindrical artificial gravity
surface by elevator tubes.

Thick walls that play the role of a radiation shield are shown in
reddish colour in Fig.~\ref{fig:schematic3D}.

\begin{figure*}[htbp]
\centering
\includegraphics[width=0.85\textwidth]{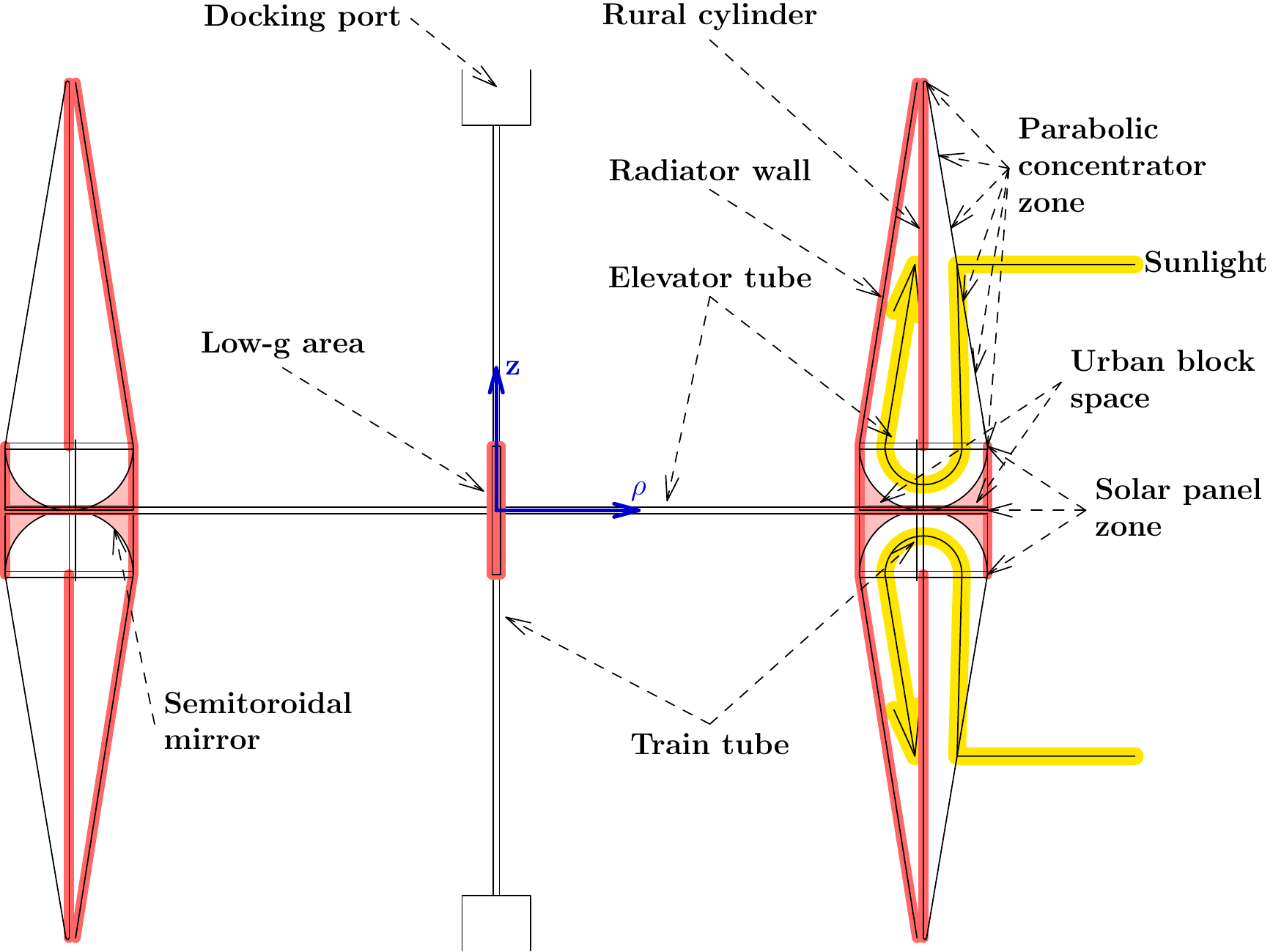}
\caption{
2-D cut view of the cylindrical space \MARKIV{settlement}. Thick radiation
shielding parts of the walls are shown in red.
}
\label{fig:schematic2D}
\end{figure*}

A 2-D cross section of the \MARKIV{settlement} is shown in
Fig.~\ref{fig:schematic2D}, where the radiation shielding parts are
marked red. The radiation shielding walls are positioned so that no
direct paths exist between the living wall surface and external space.

The main parameters of the \MARKIV{settlement} are listed in Table
\ref{tab:params}. The light transfer strategy is scalable to any
\MARKIV{settlement} size. The main underlying assumptions are summarised in Table
\ref{tab:assumptions}.

\begin{table*}[htbp]
\centering
\caption{Main parameters of the \MARKIV{settlement}.}
\begin{threeparttable}
\begin{tabular}{ll}
\hline
Diameter           & 11.5 km \\
Length             & 10.01 km\tnote{*} \\
Radiation shield wall mass & $3.55\cdot 10^{12}$ kg \\
Total mass         & $\sim 4\cdot 10^{12}$ kg \\  
Rural wall radius  & 5.0 km \\
Rural wall length  & $2 \times 4.255$ km \\
Rural area         & 267 km$^2$ \\
Number of valleys  & $2 \times 20$ \\
Valley width       & 1.57 km \\
Valley length      & 4.255 km \\
Valley area        & 6.68 km$^2$ \\
Reflector semitorus radius & 750 m \\
\MARKI{Sun-facing solar panel area} & \MARKI{17.25 km$^2$ ($11.5\,\mathrm{km} \times 1.5\,\mathrm{km}$)} \\
Electric power /w 20\,\% solar panel efficiency    & 4.7 GW \\
Artificial gravity at rural wall & 0.93 g \\
Rotation period    & 2.45 min \\
\hline
\end{tabular}
\begin{tablenotes}
  \item[*] Excluding docking ports.
\end{tablenotes}
\end{threeparttable}
\label{tab:params}
\end{table*}

\begin{table}[htbp]
\centering
\caption{Underlying assumptions.}
\begin{threeparttable}
\begin{tabular}{ll}
\hline
Solar distance                    & 1 au \\
Solar constant after filtering out IR,UV & 1000 W/m$^2$ \\
Inertia moment ratio $I_{zz}/I_{xx}$ & 1.2 \\
Mirror absorptivity               & 0.02 \\
Mirror diffuse reflectance fraction & 0.001 \\
Concentration factor, $K$ & 20 \\
Maximum rural time-average insolation & 160 W/m$^2$ \\
Minimum rural time-average insolation & 100 W/m$^2$ \\
Spatiotemporal average rural insolation & 130 W/m$^2$ \\
Rural albedo                      & 0.2 \\
Radiation shielding mass\tnote{*} & $10^4$ kg/m$^2$ \\
\hline
\end{tabular}
\begin{tablenotes}
  \item[*] Doubles as structural mass.
\end{tablenotes}
\end{threeparttable}
\label{tab:assumptions}
\end{table}

We adopt a population density of 500 persons per square kilometre of
rural \MARKIV{wall} area. For comparison, The Netherlands has a population
density of 400/km$^2$ and is a net exporter of food. It is a
requirement that a space \MARKIV{settlement} produces all the food that it consumes
in a closed ecosystem, with sufficient margin.

The \MARKIV{settlement} length is selected by requiring, following the Kalpana One
design choice, that the inertial moment tensor component $I_{zz}$ is
20\,\% larger than the $x$ and $y$ components: $I_{zz}=1.2 I_{xx}$.
This is thought to be enough to have a passively stable rotation with sufficient margin.
Here $I_{xx}=I_{yy}$ holds because of the cylindrical symmetry.
Only the radiation shielding walls were considered when
calculating the inertial moment, because they are the heaviest parts.

\begin{figure*}[htbp]
\centering
\fbox{\begin{overpic}[width=0.63\columnwidth]{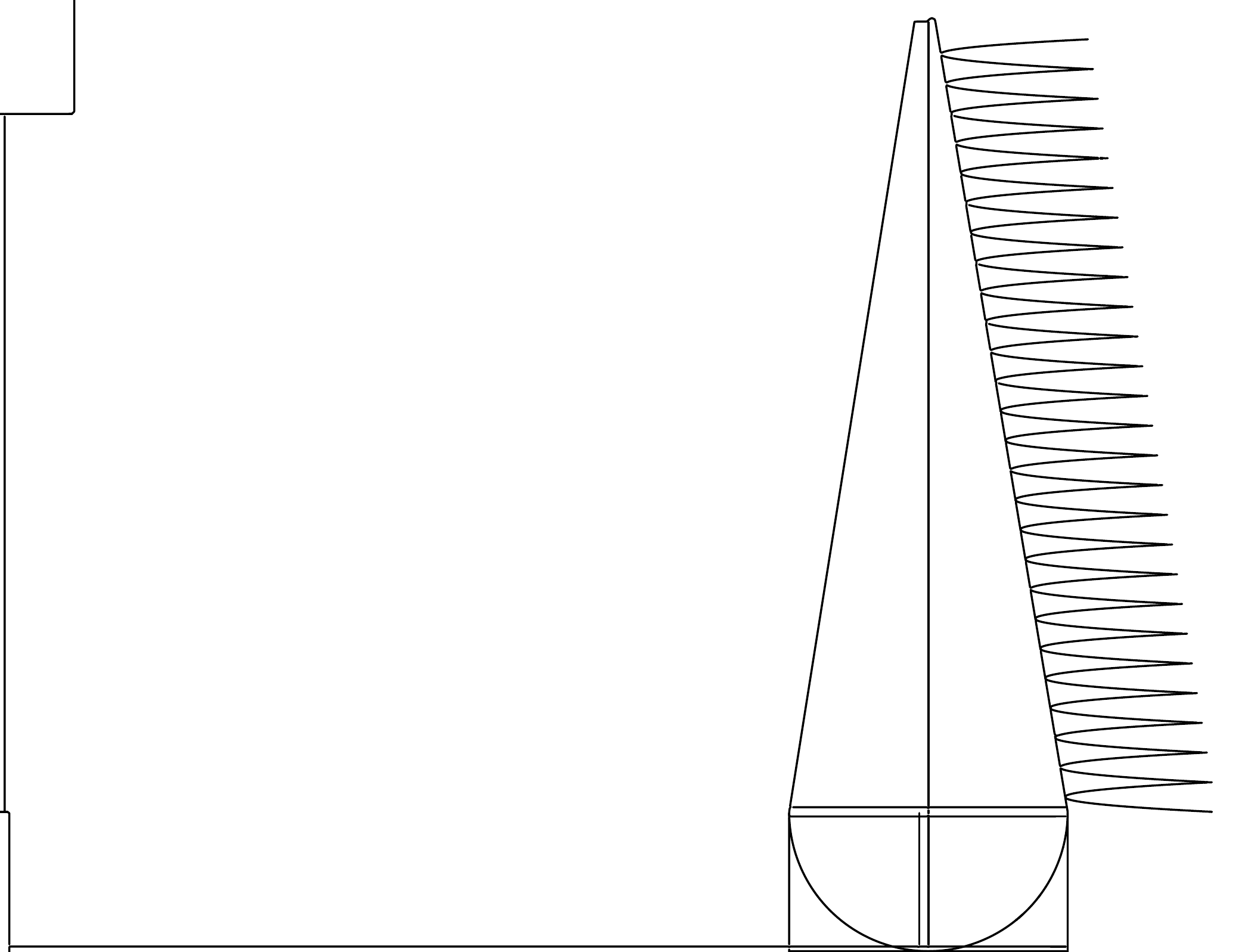}\put(0,70){a)}\end{overpic}}
\fbox{\begin{overpic}[width=0.63\columnwidth]{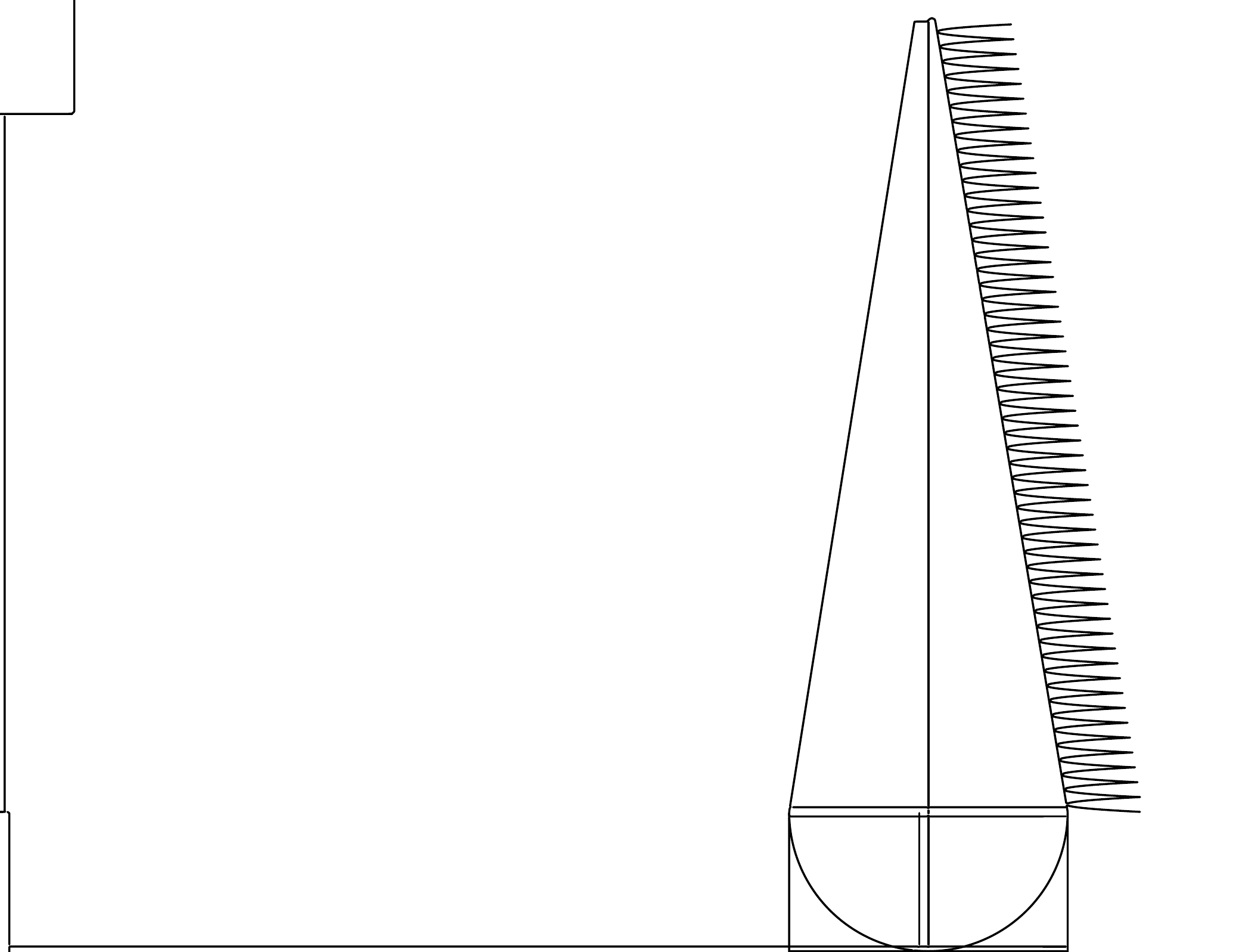}\put(0,70){b)}\end{overpic}}
\goodbreak
\fbox{\begin{overpic}[width=0.63\columnwidth]{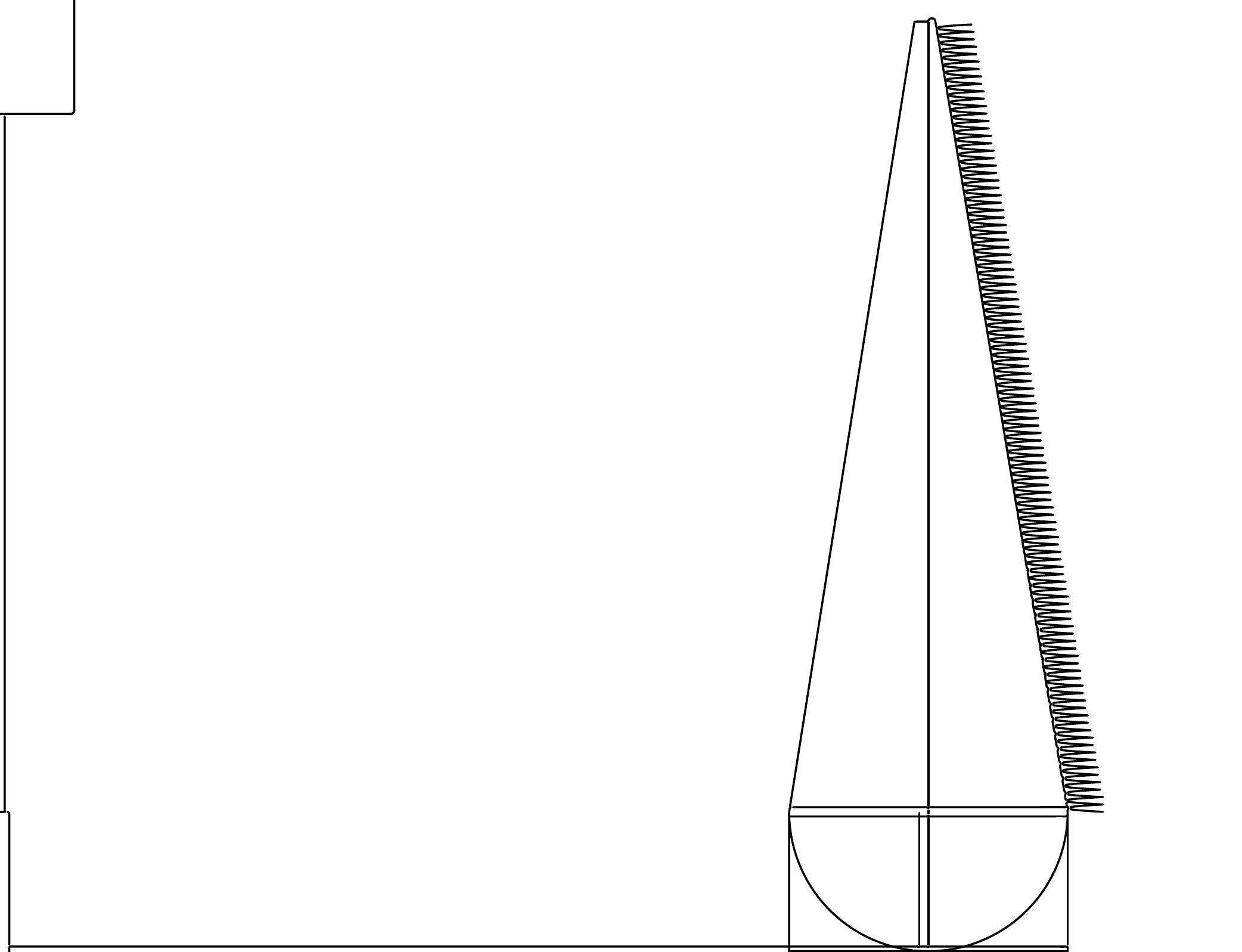}\put(0,70){c)}\end{overpic}}
\fbox{\begin{overpic}[width=0.63\columnwidth]{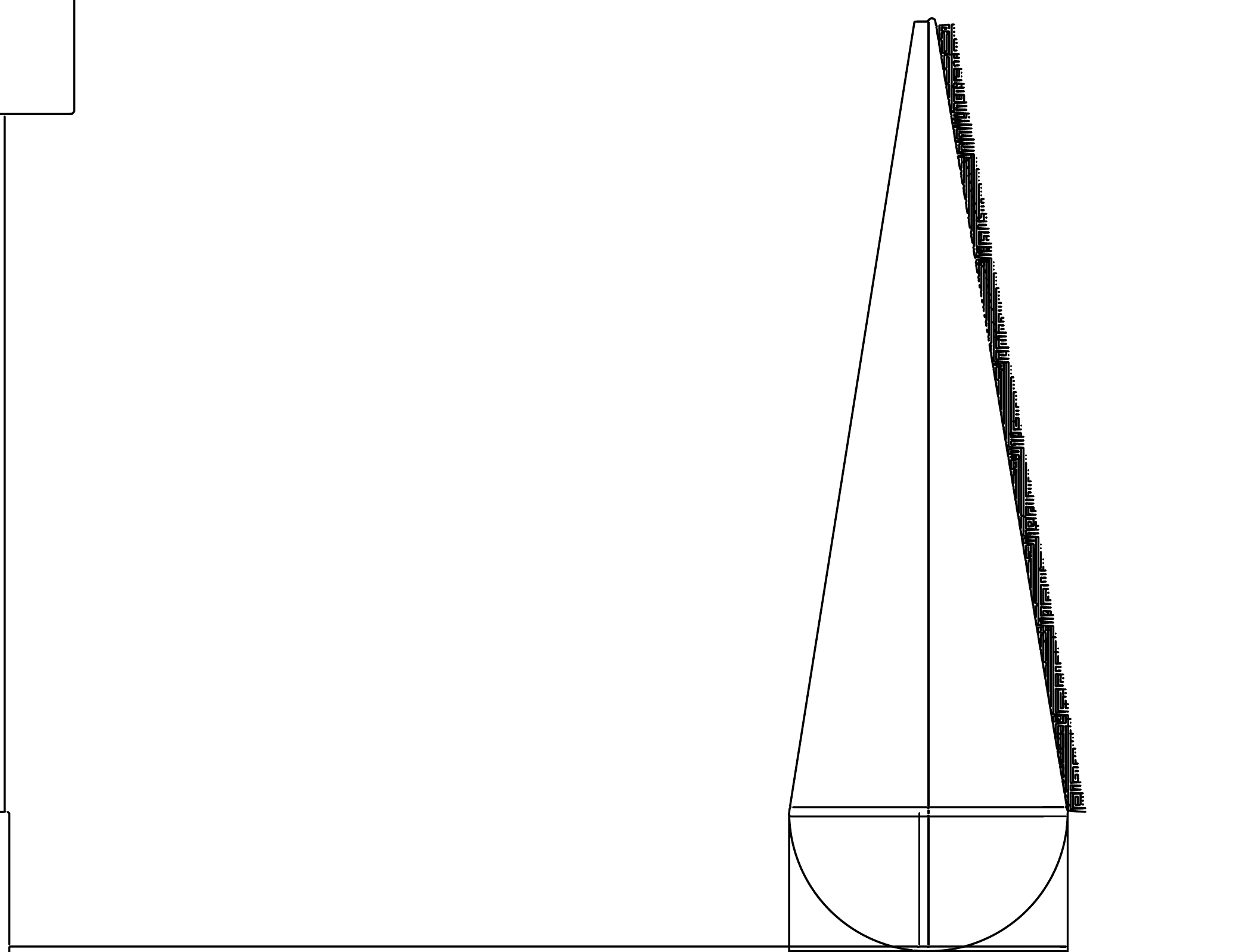}\put(0,70){d)}\end{overpic}}
\caption{
2-D cuts showing the parabolic concentrators for different slit half-width
parameters $p$ of 4 m (a), 2 m (b), 1 m (c), 0.5 m (d). The value
$p=1$ m (panel c) is used in the calculations. \MARKIV{In the limit of
  small $p$ (small and numerous concentrators), the parabolic
  concentrator zone looks dark and featureless to an
  external viewer, as was depicted in Fig.~\ref{fig:schematic3D}.}
}
\label{fig:parabolas}
\end{figure*}

Figure \ref{fig:parabolas} shows 2-D cross sections of the \MARKIV{settlement} with
the paraboloid concentrators in different scales, i.e.~different
\emph{semilatus rectum} parameters $p$. In an actual \MARKIV{settlement} one could
have a large number of small concentrators (small $p$, perhaps a few
centimetres). Simulating a small $p$ is time-consuming and in the
calculations we use $p=1$ m (Fig.~\ref{fig:parabolas}c).  We have
verified that the results are not sensitive to the value of $p$.

\section{Illumination simulation}

We wrote a \verb!C++14! \citep{cplusplus} code that performs forward and backward
ray-tracing. Figure \ref{fig:schematic3D} was produced by the code by
backward ray-tracing. Forward ray-tracing is used to launch $4\cdot
10^6$ random solar photons towards the \MARKIV{settlement}. Each photon typically
undergoes many specular and diffuse reflections until it is absorbed
by some of the optical surfaces or by the \MARKIV{settlement}'s biosphere. The
photon may also exit back to space through a paraboloid concentrator
slit. The ray-tracing code supports geometric shapes such as
planes, cylinders, cones, tori and cylindrical paraboloids, as well as
the set operations union, intersection, complement and
difference to form more complicated surfaces.

We start with a trial value of 0.08 for the biosphere's
absorptivity. The absorptivity value models both the blinders and the
biosphere itself. The resulting absorbed power density is calculated
as a gridded function of $\varphi$ and $z$ on the \MARKIV{cylindrical}
rural \MARKIV{wall}. We then calculate the corresponding insolation (power density that would
be absorbed by a small piece of fully absorbing material) by dividing
the actually absorbed power density by one minus the albedo of the
biosphere, for which we use the value 0.2. We compare the obtained
insolation with the wanted insolation which varies between 100 and 160
W/m$^2$ in $z$. Then we multiply the local absorptivity by the ratio of
the wanted versus obtained insolation and compute the next
iteration. We perform four iterations, during which the result
typically converges well. As a result, the code reproduces the wanted
insolation profile, using some absorptivity profile that depends on
$\varphi$ and $z$.

\begin{figure}[htbp]
\centering
\begin{overpic}[width=\columnwidth]{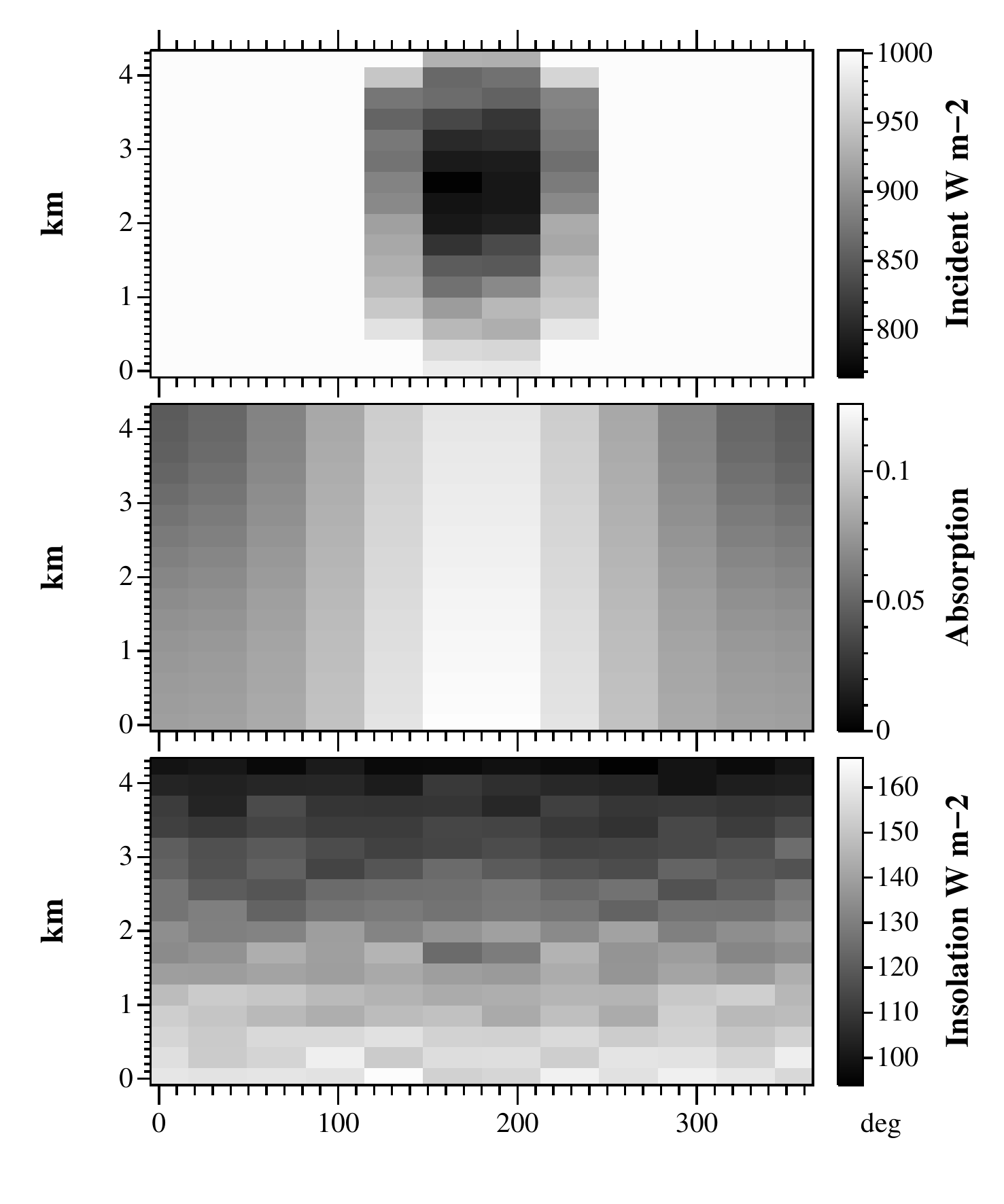}
\bfseries
\put(16,92){a)}
\color{white}
\put(16,62){b)}
\put(16,33){c)}
\end{overpic}
\caption{
Incident photon power density above the blinders (a), effective absorptivity of the blinder-equipped \MARKIV{settlement} surface (b),
insolation below the blinders. All are as function of the cylindrical coordinates $\varphi$
and $z$. In (a), values over 1 kW are shown are white.
}
\label{fig:figdens}
\end{figure}

We show the results in Fig.~\ref{fig:figdens}. Figure \ref{fig:figdens}a
shows the incident illumination above the blinders, which is
calculated by dividing the absorbed power density by the local
absorptivity. (We checked that we obtain the same result from a
transparent photon-counting detector near the surface.) The incident illumination represents the highest
obtainable illumination that results if all blinders are fully open
and the roof is completely transparent. For most $\varphi$ values the
incident illumination exceeds 1000 W/m$^2$. The value 1000 W/m$^2$
corresponds to sunshine on a cloudless day when the sun is at the zenith.
The lowest illumination that occurs for $\varphi \approx 180^{\circ}$
is 850 W/m$^2$. Each point on the biosphere passes through all $\varphi$ values in a few minutes as the
\MARKIV{settlement} rotates. Hence, for each $z$ the minimum incident flux
corresponds to the maximum obtainable illumination, if one wants to
avoid periodic dimming of the daylight at the rotational frequency.

Figure \ref{fig:figdens}b is the corresponding time-averaged absorption, which
includes the effect of the blinders as well as the biosphere
itself. The lowest panel Fig.~\ref{fig:figdens}c is the obtained
insolation below the blinders. Apart from numerical noise originating
from the finite number of photons used in the calculation, the result
corresponds to the wanted insolation profiles that goes from 160 to
100 W/m$^2$ when one moves from the equatorial end to the high
latitude end.

\MARK{

The cylindrically symmetric light channel distributes light in
$\varphi$ and $z$. Figure \ref{fig:photons} shows paths of photons
that enter the \MARKIV{settlement} in a particular exemplary $y=\mathrm{const}$
plane ($y=0.5 R$). To ease visualisation, views from two directions
are included. Some photons are reflected out already when interacting
with the cylindrical paraboloid concentrator. A few photons find their
way out from the concentrator slits later after reflecting back and
forth inside the light channel. A significant amount of photons go to
the antisunward side, some even make a full circle around the
\MARKIV{settlement}. The Monte Carlo approach was used in the simulation, i.e., at
each reflection the photon also has a chance of being absorbed at a
nonzero probability.  In Fig.~\ref{fig:photons}, the path of each
photon was traced until it escaped or was absorbed at an internal
wall. The internal walls include reflector surfaces that are part of the light
channel, blinders covering the windows between the light channel and
the rural wall, and the rural wall itself.

\begin{figure}[htbp]
\centering
\fbox{
\newcommand{\x}{50}
\newcommand{\y}{17.4}
\begin{overpic}[width=0.92\columnwidth]{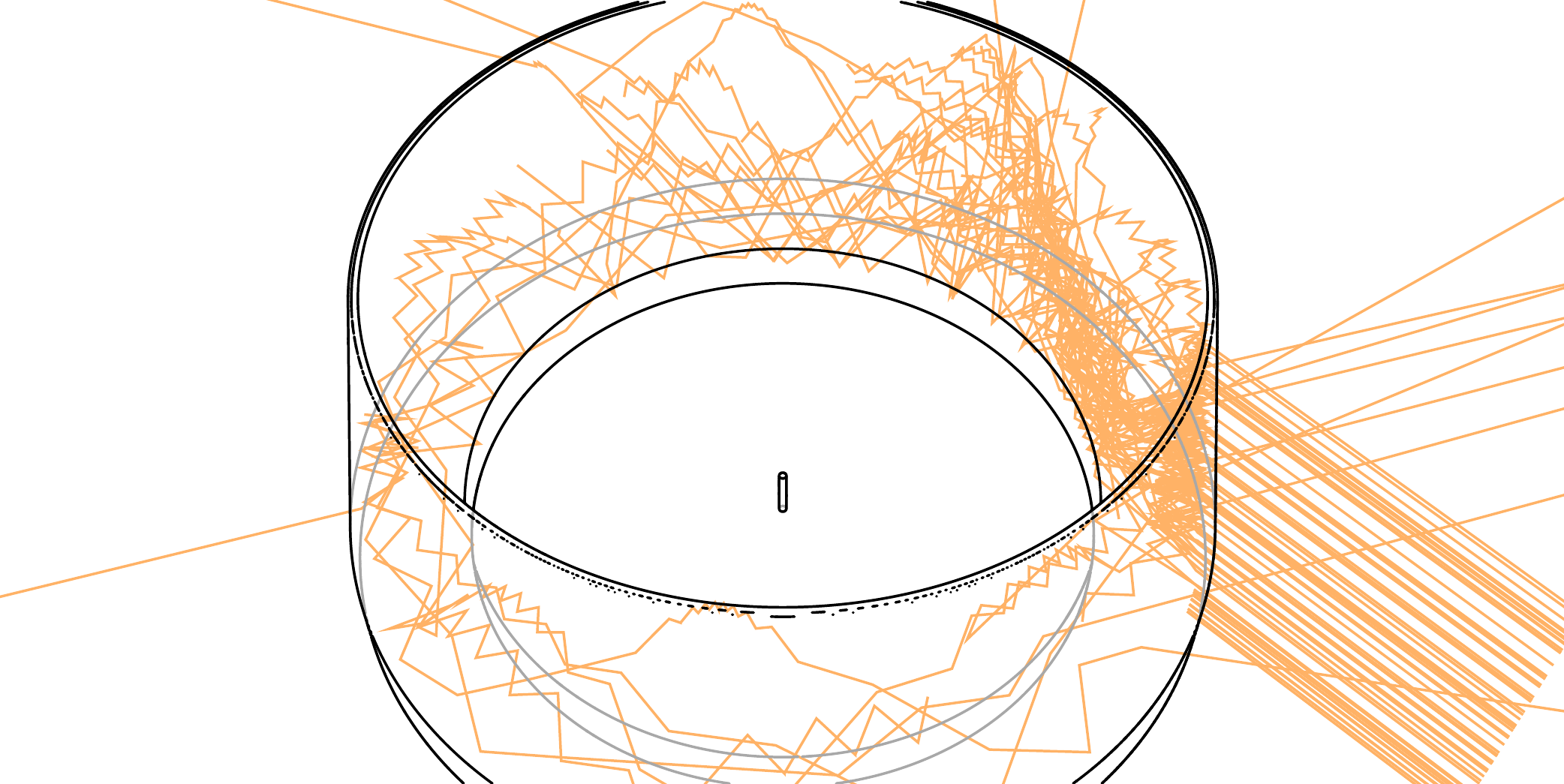}
\bfseries
\put(3,44){a)}
\put(\x,\y){\vector(0,1){8}}
\put(\x,\y){\vector(1,-0.76){7}}
\put(\x,\y){\vector(1,0.68){7}}
\put(49,26.4){Z}
\put(57,21.5){Y}
\put(56.5,12.9){X}
\end{overpic}
}
\fbox{
\newcommand{\x}{50}
\newcommand{\y}{12.8}
\begin{overpic}[width=0.92\columnwidth]{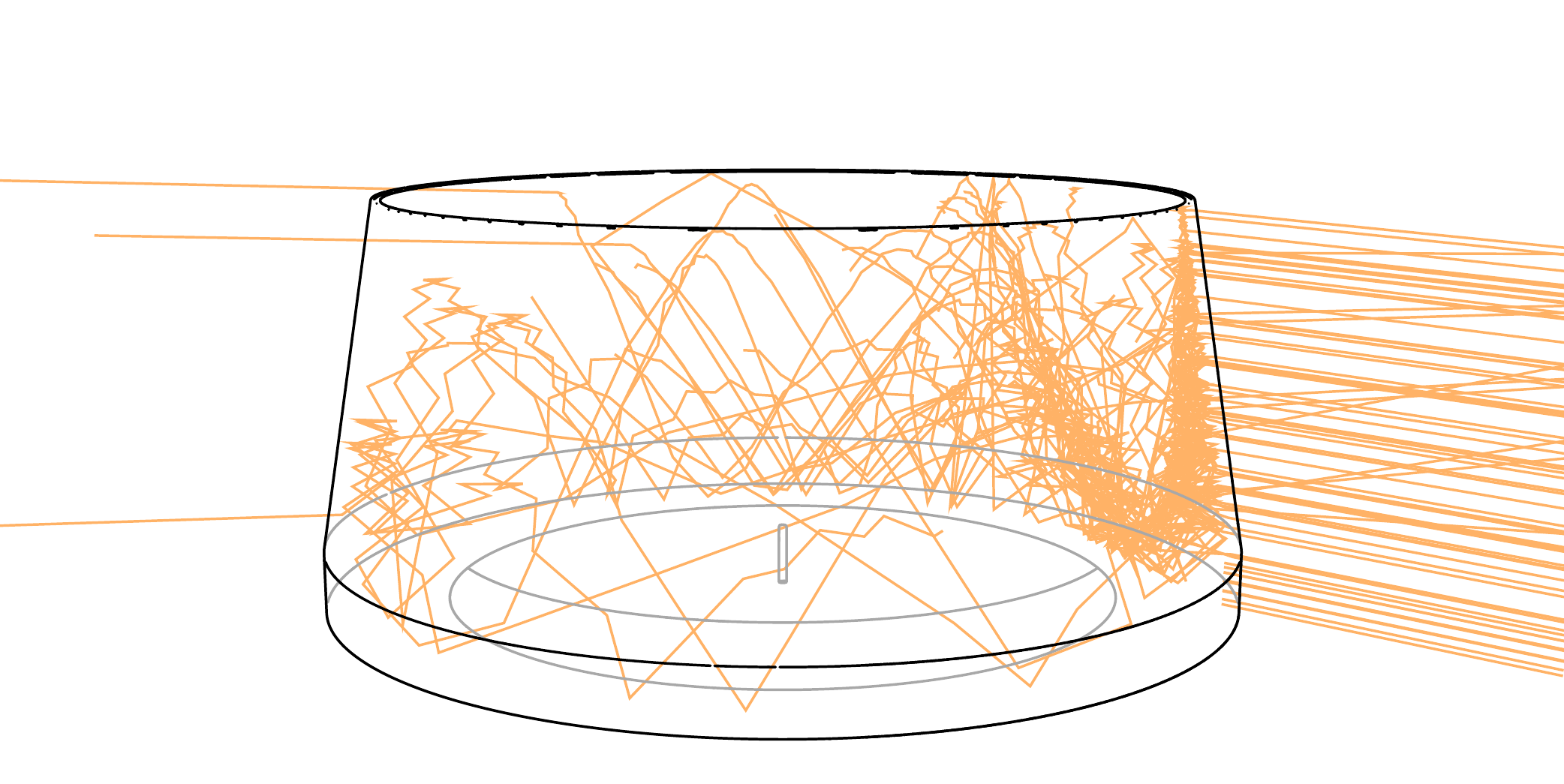}
\bfseries
\put(3,44){b)}
\put(\x,\y){\vector(0,1){8}}
\put(\x,\y){\vector(1,-0.2){7}}
\put(\x,\y){\vector(1,0.25){7}}
\put(49,21.2){Z}
\put(57,14){Y}
\put(57,10.4){X}
\end{overpic}
}
\caption{\MARK{Photon paths inside the light channel from two viewing
angles. A vertical sheet of input photons enters from the right
hand side and propagates along $-\hat{x}$ at $y=R/2$, while $z$ is a
random number in interval $\left[0,Lz/2\right)$.}}
\label{fig:photons}
\end{figure}
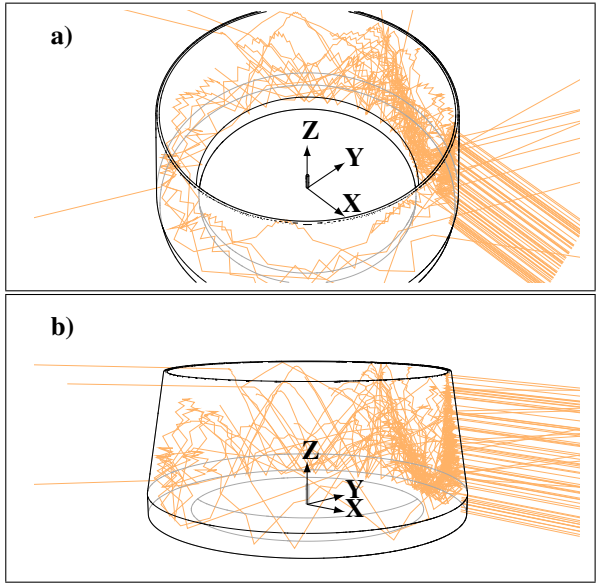

}  

\begin{figure}[htbp]
\centering
\includegraphics[width=\columnwidth]{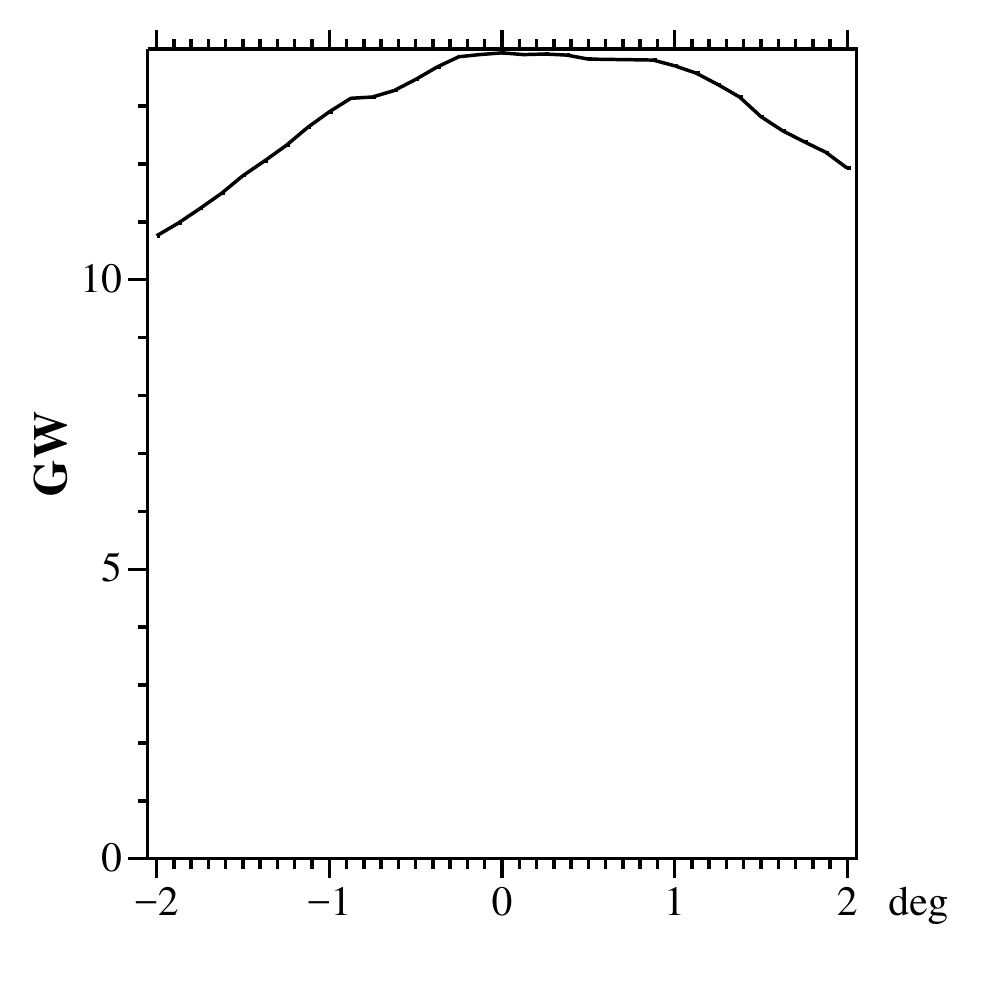}
\caption{
Power absorbed by the living cylinder as function of sun angle.
}
\label{fig:sunangle}
\end{figure}

The cylindrical paraboloid concentrators are somewhat sensitive to
the direction of the incoming sunlight. Increasing the concentration
ratio $K$ (i.e.~the total width versus slit width of the parabola)
traps light better because it reduces the amount of light that escapes
through the slits. On the other hand, increasing $K$ increases the
directional sensitivity. We use $K=20$ as a tradeoff. Figure
\ref{fig:sunangle} shows the total sunlight power absorbed by the
rural biosphere (in the $z>0$ half of the \MARKIV{settlement}) as a function of the
angle between the sun direction and the spin plane. We see in
Fig.~\ref{fig:sunangle} that if the spin axis is controlled with
better than $\sim 0.5^{\circ}$ precision, then the absorbed power is
optimal, while a deviation of 2 degrees causes up to 20\,\%
reduction. Even a 20\,\% reduction is not catastrophic for the crop
yields, however, since plants are accustomed to cloudiness associated
insolation variations on Earth during the growing season. The temperature remains
controllable by regulating the cooling through the blinders and
the radiator wall. We think that the sensitivity to spin axis
inclination is not a problem. The sun angle dependence
(Fig.~\ref{fig:sunangle}) is not quite symmetrical because although
the southern and northern parts of the \MARKIV{settlement} are identical, there is
no light exchange between them and the northern part taken alone is
not symmetrical in $z$.

The \MARKIV{settlement} is cooled radiatively through the inner conical reflector
wall whose interior emits thermal infrared into space. A glass-covered
optical mirror of the inner side of the conical surface absorbs well
in thermal infrared and conducts heat away because its underlying
support structure is metallic. To avoid using heat pumps, the radiator
temperature must be lower than the dewpoint of the climate that one
wants to maintain. In our simulation, the average radiator temperature
is $-16$ $^{\circ}$C if one makes the approximation that the effective
radiating areas are the open endcap areas of the cylinder
\MARKIII{(making such assumption models the effect that part of the
  emitted infrared is re-absorbed by the opposing radiator wall)}. Power
generated at the rural \MARKIV{wall} is radiated away by the nearby
radiator wall so that liquid transfer of heat over longer distances is
not required.

\section{Urban blocks}

The population density per rural wall area is limited by the
biological productivity of the rural walls, we have assumed 500
persons per square kilometre, or 2000 m$^2$ of rural area per
person. Additional space per person can be obtained by building urban
type multi-storey blocks, and it can be done without reducing the
rural area. For example, urban blocks could simply exist on the rural
walls so that the roofs of the buildings are used for
agriculture. Another possibility which is better from the mass
distribution point of view is to build urban blocks inside the volume
(Fig.~\ref{fig:schematic2D}, pale red areas) that is left over by the
equatorial semi-toroidal reflectors.  Equatorial mass promotes
rotational stability, and the extra stability would enable making the
\MARKIV{settlement} somewhat longer, with a corresponding increase in rural area
and consequently the maximum population.

Table \ref{tab:population} lists population related parameters. For a
\MARKIV{settlement} of 5 km rural wall radius, the space available for urban blocks
between the toroidal reflectors (pale red areas in
Fig.~\ref{fig:schematic2D}) is as large as 15 km$^3$. The volume
scales cubically with the \MARKIV{settlement} radius. The volume corresponds to a
vast $25,000$ m$^2$ of urban floor area per person, which is an order
of magnitude larger than the rural area per person. In this paper we
call this the maximal (or naturally buildable maximal) urban area,
even though it is not really maximum because one could build also
additional urban spaces in the rural areas.

To have a rough estimate of the mass of urban blocks, the large luxury
cruise ship ``Oasis of the Seas'' weighs $10^8$ kg and its total
internal volume is $710.6\cdot 10^3$ m$^3$, as calculated from the
gross tonnage of 225,282 GT\citep{OasisOfTheSeas}, yielding a bulk
density of 141 kg/m$^3$. If this bulk density is representative of the
\MARKIV{settlement} blocks as well, the mass of the fully built urban blocks
filling 15 km$^3$ volume is $2\cdot 10^{12}$ kg, which is 50\,\% of
the mass of the rural-only version of the \MARKIV{settlement} (Table
\ref{tab:params}). If one chooses to build 10\,\% of the maximum urban
volume, for example, it increases the \MARKIV{settlement} mass by only 5\,\% and provides
2000 m$^2$ of urban floor area per person, which is
still a vast amount.

\begin{table}[htbp]
\centering
\caption{Population parameters.}
\begin{threeparttable}
\begin{tabular}{ll}
\hline
Total population   & 134,000 \\
Population density per rural area & 500 km$^{-2}$ \\
Rural area per capita & 2000 m$^2$ \\
Electric power per capita   & 35 kW \\
Max equatorial urban block volume & 15 km$^3$ \\
Max urban floor area \tnote{*} & $3.3\cdot 10^9$ m$^2$ \\
Max urban volume per capita & $110,000$ m$^3$ \\
Max urban floor area per capita & $25,000$ m$^2$ \\
El.~power per fully built urban area & $1.5$ W/m$^2$ \\
Artificial gravity in urban blocks & 0.79--1.07 g \\
\hline
\end{tabular}
\begin{tablenotes}
  \item[*] Assuming 4.5\,m floor-to-floor height difference.
\end{tablenotes}
\end{threeparttable}
\label{tab:population}
\end{table}

The available electric power is large per capita (35 kW), although
lowish per maximal urban floor area (1.5 W). The artificial gravity in
the inner and and outer urban volume is lower and higher,
respectively, than on the rural walls. In Tables \ref{tab:params} and
\ref{tab:population} we have selected the $1 g$ level to be reached
halfway between the rural wall and the outermost part of the urban
blocks. For example, kindergartens might exist in the $\ge 1 g$ parts
so that children grow as strong as on Earth, whereas motion-challenged
old people might live in the parts were the gravity is lower than $1 g$.

\MARKIII{The interiors of the urban blocks should also be shielded against
  radiation. We do not include mass for such shielding in the mass
  budget explicitly, because
  we consider it likely that parts of the urban blocks are used as storage
  area for items that do not need radiation protection. By placing the storage
  areas at the outer boundary of the urban block, their mass contributes to the
  shielding of the inside of the block so that dedicated radiation shielding
  might be unnecessary.}

\section{Discussion}

We assumed a 5 km \MARKIV{settlement} radius. The value is meant to correspond
roughly to the sweet spot where the structural walls double as
radiation shields. \MARKIV{Settlements} of all sizes up to and including the sweet
spot size have the same wall thickness and thus the same mass cost per
inhabitant\footnote{The sweet spot size is a function of the radiation
  environment. Thus an equatorial LEO \MARKIV{settlement} would have a much smaller
  sweet spot size than the deep space \MARKIV{settlement} considered in this
  paper.}. The sweet spot size is preferred because a large \MARKIV{settlement} is
preferable over a small one. \MARKIV{Settlements} larger than the sweet spot size
have, however, a higher mass cost per inhabitant because of
structural requirements. Because of this, we consider it likely that
large-scale \MARKIV{settling} of the solar system is undertaken using
\MARKIV{settlements} around the sweet spot size.

We assumed light channel width which is 15\,\% of the rural wall
radius (750 m, since the rural wall radius is 5 km). A wider light
channel would distribute sunlight more evenly, but would increase the
mass needed per illuminated rural area because a larger fraction of
the \MARKIV{settlement} length would be radiation shielding surface.

Periodic variation of the insolation by the rotation rate is not
observable by the inhabitants because it is suppressed by continuous
active controlling of the blinders. The diurnal and seasonal cycles
are produced synthetically by the blinders. Failure of a small
fraction of the blinders is acceptable. The blinders can be lubricated
in a normal manner because they reside in a 1 bar pressure normal
atmosphere in normal temperature. They can be serviced and replaced by
humans and robots easily because they reside in earthlike gravity and
radiation protection.

We divided the \MARKIV{cylindrical} rural \MARKIV{wall} into (for example) 20 valleys in the
$\varphi$ direction. Each valley is in its own phase of the diurnal
and seasonal cycle. The arrangement enables simulating the diurnal and
seasonal cycles using local blinders while keeping the total sunlight
power dissipation roughly constant. Neighbouring valleys are isolated
from each other optically, and for safety reasons they can also be
pressurised independently of each other. Division of the rural
\MARKIV{wall} cylinder into valleys also has the benefit of making the cylinder's
curvature visually less apparent to the inhabitants, particularly if
there are artificial ridges or ``mountains'' between the valleys.

One benefit of the proposed architecture is that thick radiation and
meteoroid protecting transparent windows are not needed.

It is noteworthy that the naturally available maximal urban floor area
per inhabitant is large, 25,000 m$^2$ under the baseline
assumptions. This floor area per person is 2-3 orders of magnitude
larger than the contemporary average on Earth and comparable to what
today's royal families have access to. If the urban blocks are maximally
built, electric power usage per urban floor area must be limited to
1.5 W/m$^2$. However, this does not restrict the use of the floorspace
in an essential way because rooms where no people are present do not
need energy. Plants feeding the biosphere must be grown in the open rural
areas where the light energy dissipation per area is two orders of
magnitude higher. The dominant power is the sunlight dissipated in the
rural areas, while for a maximally built \MARKIV{settlement} the dominant floor
area exists in electrically lighted urban blocks.

The $3.3\cdot 10^9$ m$^2$ urban floor area is so large that if
arranged in rooms of 6.7 m wide, for example, a path going through all
the rooms must be at least $5\cdot 10^5$ km long. A person walking 20
km per day for 70 years could theoretically visit every room
once. Thus, a 5 km radius \MARKIV{settlement} can have more things inside than an
inhabitant can experience or explore over an entire lifetime. Even so,
the maximal urban volume is only $\sim 6$\,\% of the volume of the
light channels.

We calculated the length of the \MARKIV{settlement} from the requirement
$I_{zz}=1.2\cdot I_{xx}$, without including the mass of any urban
blocks. If one decides to introduce large urban blocks, the \MARKIV{settlement}
could be made somewhat longer while keeping passively stable
rotation. It is also possible and perhaps likely that urban blocks are
constructed gradually by the first generations of inhabitants. In that
case, the \MARKIV{settlement} length must be restricted so that rotation is stable
also without the mass of the urban blocks, which is the case that was
calculated in this paper.

\MARKIV{The inner and outer light channels are tapered,
  i.e.~the channel gets
  narrower with increasing $\vert z\vert$ as seen e.g.~in Fig.~\ref{fig:schematic2D}.
  A tapered inner light channel works
  better than a non-tapered one, for multiple reasons:
\begin{enumerate}
\item According to numerical experimentation, a tapered inner light channel
  distributes light better onto the rural wall. The sunlight power that
  reaches the rural wall is higher and more uniform in $\varphi$ and $z$.
\item A non-tapered inner light channel would need to be closed by a
  radiation shielding end member.
  The radiation shield would increase the settlement's total mass, and
  the extra mass would be located at high $\vert z\vert$ which would reduce the
  stability of the rotation. To restore stable rotation, the cylindrical
  rural wall would have to be made shorter, which would result in smaller area for the
  food-producing biosphere and thus a smaller maximum population.
\item Cooling of the settlement occurs by thermal radiation emitted by
  the inner wall of the inner light channel\footnote{\MARKIV{In this cooling method,
  heat produced at the rural wall by dissipated sunlight is
  radiated away by the adjacent inner lightchannel surface so that
  there is not much if any need for fluid-based heat transfer.}}. The emitted thermal radiation is
  partially re-absorbed by the opposing wall. To account for the re-absorption,
  the effective radiator area is approximately given by the
  area of the open end of the cylinder. The open area is larger if the
  inner light channel is tapered. Hence, tapering increases the
  available cooling power of the settlement, which increases
  the maximum tolerable sunlight power that the rural wall can be configured to absorb by the
  artificial diurnal and seasonal illumination cycles. Consequently,
  when all else is equal, the maximum biological production and
  the maximum allowed human population are larger when
  the inner light channel is tapered.
\end{enumerate}

Tapering is beneficial also for the outer light channel because
according to numerical experimentation, a tapered channel directs light towards the
equatorial semitoroidal reflector with fewer reflections, so that less
light gets absorbed by the reflector surfaces or escapes through the
concentrator slits back into space.
}

Are there other ways to design a \MARKIV{settlement} that produces earthlike
illumination cycles without large-scale moving parts? One approach
would be to use artificial light sources instead of natural sunlight
and to produce the needed electric power by covering the exterior
surface by solar panels. However, natural sunlight is expected to be
visually more pleasing than artificial light and it also has cost,
lifetime, reliability and thermal management benefits relative to
electric lighting.

\section{Summary and conclusions}

We have presented a cylindrical space \MARKIV{settlement} concept where sunlight is
concentrated by cylindrical paraboloid concentrators and reflected by
semi-toroidal and conical reflectors and controlled by local blinders
to simulate earthlike diurnal and seasonal illumination cycles. The
rural \MARKIV{wall} living cylinder is divided into 20 (for example) $z$-directed
valleys which are in different phases of the light cycles. No moving
parts are needed other than numerous and easily accessible local
blinders that regulate light input into the valleys. The \MARKIV{settlement}
rotates as a rigid body and the mass distribution is such that the
rotation is passively stable.

The geometry has natural spare volume at the equator where one can add
multi-storey urban blocks without reducing the rural area.  Adjacent
to the urban blocks there is natural place where to install a zone of
solar panels without reducing the amount of gathered light.

The inhabitant population is limited by the ability of the closed ecosystem
of the sunlit rural areas to produce food. The naturally buildable total urban floorspace
area can be an order of magnitude larger than the rural area if the
\MARKIV{settlement} radius is 5 km. The maximum urban to rural area ratio scales
linearly with the \MARKIV{settlement} radius.  A 5 km \MARKIV{settlement} radius corresponds roughly
to the sweet design spot where earthlike radiation shielding is
produced for free by the required structural mass.

Overall, the \MARKIV{settlement} concepts satisfies the following generic requirements for
long-term large-scale \MARKIV{settling} of the solar system:
\begin{enumerate}
\item $1 g$ artificial gravity, earthlike atmosphere, earthlike
  radiation protection.
\item Large enough size so that internals of the \MARKIV{settlement} exceed a
  person's lifetime-integrated capacity to explore.
\item Standard of living reminiscent to contemporary royal families on
  Earth, quantified by up to 25,000 m$^2$ of urban living area and 2000 m$^2$
  of rural area per inhabitant.
\item Access to other \MARKIV{settlements} and Earth by spacecraft docking ports,
  using safe arrival and departure procedures that do not require impulsive
  chemical propulsion.
\end{enumerate}

In particular, the proposed lighting geometry enables a long-term
reliable architecture, i.e., a design that does not include large
moving parts, is free of single failure points and exhibits passively
stable rotation.

As a future refinement of the concept, one could consider more general
orbits than 1 au circular, \MARKII{and a range of \MARKIV{settlement} sizes could
  be analysed}.

In conclusion, a requirement for \MARKIV{settling} the solar system in a
large scale is that the habitats must be long-term reliable and they
should provide a high standard of living compared to Earth. In the
light of our analysis, the goals seem possible to reach, without
essentially increasing the total mass consumption per inhabitant beyond what
is required by radiation protection in any case.

\section{Acknowledgement}

The results presented have been achieved under the framework of the
Finnish Centre of Excellence in Research of Sustainable Space (Academy
of Finland grant number 312356). \MARK{The author is grateful to Sini
  Merikallio for suggesting a modification of the title. The author
  also thanks \MARKIV{the Editor Al Globus and} three
  \MARKIV{anonymous reviewers} for making a large number of very
  useful constructive comments.}

\appendix

\MARKI{

\section{Wall tension}


Consider a cylindrical \MARKIV{settlement} wall of radius $R$, length $L_z$,
thickness $h$ and wall mass density $\rho_w$, rotating so that the
centrifugal acceleration at the wall is $g$. The mass of the cylinder
is
\begin{equation}
m = 2\pi R L_z h \rho_w.
\end{equation}
Use cylindrical coordinates $\rho$, $\varphi$ and $z$ so that the
cylinder is centred at origin and its axis is aligned with
$\hat{z}$. Let us consider the $y>0$ half of the cylinder.
For this semicylinder $0\le\varphi<\pi$.

The wall tension $F_y$ at $y=0$ is equal to the $y$-component of the
centrifugal force acting on the $\{y>0\}$ semicylinder:
\begin{equation}
F_y = \int dF_y = \int dm g \sin\varphi = \frac{m g}{2\pi}
\int_0^\pi d\varphi \sin\varphi = \frac{m g}{\pi}.
\end{equation}
The force $F_y$ pulls a cross-sectional wall area
\begin{equation}
A_\perp = 2 L_z h
\end{equation}
where factor 2 comes from two edges of the semicylinder at $x=\pm R$.
The tensile stress $\sigma$ of the wall is then given by
\begin{equation}
\sigma = \frac{F_y}{A_\perp} = R\rho_w g.
\label{eq:sigma}
\end{equation}
If $R=5$ km, $\rho_w=7.8\cdot 10^3$ kg/m$^3$ (steel) and $g=9.81$ m/s$^2$, we
obtain tensile stress $\sigma=380$ MPa. In the form of piano wire,
steel's ultimate tensile strength can exceed 2500 MPa
\citep{pianowire}. If half of the wall mass is payload (soil,
vegetation, etc.), for example, then the structural parts must have
$2\times 380=760$ MPa of usable tensile strength. This is 30\,\% of
the piano steel wire ultimate strength of 2500 MPa. In the light of
these numbers, a living wall radius of 5 km seems
feasible. Steel is a reasonable choice for the main structural material since
iron is one of the dominant elements on asteroids.

It is of interest to notice that the wall stress (\ref{eq:sigma}) does
not depend on the wall thickness $h$. This is because we did not
include the internal atmospheric pressure of the \MARKIV{settlement} in the
calculation. Neglecting the internal pressure is a good approximation
since the 50 m high atmosphere contributes only 65
kg/m$^2$ to the wall mass loading.

When the \MARKIV{settlement} radius $R$ is increased, the wall tensile strength
requirement $\sigma$ increases proportionally, according to
Eq.~(\ref{eq:sigma}). If the structural material characteristics
remain the same, this means that for a larger \MARKIV{settlement}, a smaller
fraction of the wall mass can be payload (soil etc.). If the payload
mass per area is kept unchanged, this implies that the wall's total
mass per area must increase. The question of what is the sweet spot
\MARKIV{settlement} size (i.e., size where radiation shielding and structural
requirements meet) depends on the required payload mass per living
wall area. In conclusion, the adopted 5 km \MARKIV{settlement} radius is compatible
with, for example, 50\,\% wall payload fraction and tensile loading which is 30\,\% of piano wire
steel's ultimate tensile strength.

}






\end{document}